\documentclass[usenatbib]{mnras}
\usepackage{amsmath,amsfonts,amssymb,graphicx,natbib,float,wrapfig}
\newcommand{\pks}{PKS\,B1830$-$211 }
\newcommand{\pksn}{PKS\,B1830$-$211}
\title[Methanol absorption in \pks]{Methanol absorption in \pks at milliarcsecond scales}
\author[Marshall et al.]{M.\ A. Marshall$^{1}$, S.\ P. Ellingsen$^{1}$, J.\ E.\ J. Lovell${^1}$, J.\ M. Dickey${^1}$, M.\ A. Voronkov$^{2}$, \newauthor S.\ L. Breen$^{2,3}$ \\
\\
  $^1$ School of Physical Sciences, University of Tasmania, Private Bag 37, Hobart, Tasmania 7001, Australia\\
  $^2$ CSIRO Astronomy and Space Science, Australia Telescope National Facility, PO Box 76, Epping, NSW 1710, Australia\\
  $^3$ School of Physics, University of Sydney, Sydney, NSW 2006, Australia}

\begin{document}

\label{firstpage}
\pagerange{\pageref{firstpage}--\pageref{lastpage}}
\maketitle

\begin{abstract}
Observations of the frequencies of different rotational transitions of the methanol molecule have provided the most sensitive probe to date for changes in the proton-to-electron mass ratio, $\mu$, over space and time. Using methanol absorption detected in the gravitational lens system \pksn, changes in $\mu$ over the last 7.5 billion years have been constrained to $|\Delta\mu/\mu| \lesssim 1.1 \times 10^{-7}$.  Molecular absorption systems at cosmological distances present the best opportunity for constraining or measuring changes in the fundamental constants of physics over time, however, we are now at the stage where potential differences in the morphology of the absorbing systems and the background source, combined with their temporal evolution, provide the major source of uncertainty in some systems.  Here we present the first milliarcsecond resolution observations of the molecular absorption system towards \pksn.  We have imaged the absorption from the 12.2-GHz transition of methanol (which is redshifted to 6.45 GHz) toward the southwestern component and show that it is possibly offset from the peak of the continuum emission and partially resolved on milliarcsecond scales.  Future observations of other methanol transitions with similar angular resolution offer the best prospects for reducing systematic errors in investigations of possible changes in the proton-to-electron mass ratio on cosmological scales.
\end{abstract}

\begin{keywords}
ISM -- ISM: molecules -- quasars: absorption lines -- quasars: individual (PKS 1830-211)
\end{keywords}

\section{Introduction}
The search for changes in fundamental constants is an important area of study in modern physics, providing a test for current physical theories. Fundamental constants, and dimensionless combinations of these constants such as the proton-to-electron mass ratio, $\mu=m_p/m_e=1836.1526739$,\footnote{National Institute of Standards and Technology 2014 value} are values which cannot be explained nor computed by the theories themselves. The currently accepted theory of particle physics, the standard model, does not predict variations in these constants over time or space. Modern theories about the mechanism behind dark energy, however, differ in their predictions about such variations. The theory that dark energy can be explained by the cosmological constant, introduced in the theory of general relativity, predicts no changes in constants over time and space, but an alternative rolling scalar field theory predicts that constants such as $\mu$ change with time \citep{Thom2012}. More accurate measurements that are sensitive to potential changes in fundamental constants over space or time will therefore allow useful tests and constraints on these theories. A comprehensive review of the topic is given by \citet{Uzan}.

$\mu$, the fine structure constant $\alpha=\frac{1}{4 \pi \epsilon_0} \frac{e^2}{\hbar c}$ and the nuclear g-factor $g$ are common dimensionless combinations of fundamental constants that are being investigated for spatial and temporal changes. A change in one or more of these constants can be detected through a comparison of the rest frequencies, $\nu$, of atomic and molecular transitions over time and space.  The dependence of a particular transition frequency on changes can be expressed using the relationship:
\begin{equation}
\frac{\Delta\nu}{\nu}=K_\alpha\frac{\Delta\alpha}{\alpha}+K_\mu\frac{\Delta\mu}{\mu}+K_g\frac{\Delta g}{g}
\end{equation}
where $K_{\alpha}$, $K_{\mu}$ and $K_{g}$ are sensitivity coefficients, describing how sensitive an individual transition is to variations in the corresponding dimensionless constant.  So by observing certain atomic and molecular transitions that are sensitive to these constants in different regions of space or time, we can detect changes, or lack thereof, in the fundamental constants (provided the transitions compared have different sensitivities). Transitions with large differences in their sensitivity coefficients provide the best probe into variations in these constants, with a small change in the fundamental constant causing a frequency difference that is more easily able to be detected. Pure rotational, vibrational or electronic transitions have $K_\mu = -1$, $\frac{1}{2}$ and 0 respectively \citep{Varsha}. In contrast, some observed transitions of the methanol molecule have been found to have very high sensitivity coefficients, of up to $K_\mu=-42$ \citep{Jansen}.

The $2_0 \rightarrow 3_{-1} E$ transition of methanol which has a rest frequency of 12.178 GHz is the second strongest methanol maser transition from star formation regions and has a sensitivity coefficient $K_\mu = -33$ \citep{Jansen,LevSept}.  It is also commonly observed to show absorption from cold molecular clouds \citep[e.g.][]{Peng1992}. The combination of high sensitivity coefficients and relatively high abundance in astrophysical environments, make methanol a good target species to search for a variation in $\mu$ over cosmological time scales. 

Observations of methanol at cosmological distances provide a probe which is highly sensitive to possible changes in $\mu$.
However, so far the absorbing galaxy towards \pksn, is the only region in the distant universe where methanol transitions have been detected \citep{Muller2011,Simon2012}. 
\pks is a gravitational lens system \citep{Jauncey}, containing a blazar at a redshift of $z = 2.507$ \citep{Lidman}, a primary lensing galaxy at a redshift of $z = 0.88582$ \citep{Wiklind} and potentially a second galaxy along the same line of sight at $z = 0.19$ \citep{Lovell}. The lensing galaxy is a nearly face-on type Sb or Sc spiral, which at $z = 0.88582$ corresponds to a look-back time of $\sim 7.5$ Gyr \citep[adopting a standard $\Lambda$-cosmology with $\textrm{H}_0 = 67$ km~s$^{-1}$ Mpc$^{-1}$, $\Omega_m = 0.315$ and $\Omega_\Lambda = 0.685$;][]{Planck}. The blazar is radio-loud and time variable, with a steep spectrum jet and an optically thick core component. It is gravitationally lensed by the primary lensing galaxy, to be seen as two main `hotspots', or source components. The jet of the \pks quasar forms a full Einstein ring in the radio, but the ring is only prominent at low frequencies \citep[$\lesssim 10$ GHz;][]{Jauncey}.

The two main lensed images of the source component of the blazar are separated by $\sim 1"$ on the sky, with the northeastern component stronger than the southwestern component by a factor of 1.52 \citep{Lovell+98}. \pks is one of the best known targets for studying molecular gas in absorption at intermediate redshift, with more than 40 molecular species detected in the lensing galaxy \citep[e.g.][]{Muller2014}. Molecular absorption is mostly seen towards the southwestern component. This is likely to be due to the relative placements of the sources and the lensing galaxy, with the southwestern component located behind a spiral arm of the lensing galaxy, at a distance of roughly 2 kpc from the galactic centre, whilst the northeastern source is $\sim 4$ kpc from the centre \citep{Muller2006}. 

The redshift of the absorbing system towards \pksn, is such that the absorption lines are shifted in frequency to approximately half of the rest frequency of the transition, which is used to identify them; for example, the 12.2-GHz transition that is of interest to this study is observed redshifted to $\sim 6.45$ GHz.  \citet{Simon2012} compared the 12.2-GHz and 60.5-GHz ($1_0 \rightarrow 2_{-1} E$) methanol transitions in \pks to obtain a constraint of $\Delta\mu/\mu = (0.8 \pm 2.1) \times 10^{-7}$ or $\Delta\mu/\mu < 6.3 \times 10^{-7} (3\sigma)$. \citet{Bagd_2013b} used the same transitions and also the 48.372 and 48.377-GHz methanol transitions ($1_0 \rightarrow 0_{0} E$ and $1_0 \rightarrow 0_{0} A^{+}$, respectively) to obtain a tighter constraint of $\Delta\mu/\mu = (0.0 \pm 1.0) \times 10^{-7}$. \citet{Bagd_2013a} compared ten methanol transitions from 12.2 GHz to 492.3 GHz, which resulted in $\Delta\mu/\mu = (1.5 \pm 1.5) \times 10^{-7}$. In the most stringent of constraints, \citet{Kanekar} obtained $\Delta\mu/\mu \leq 1.1 \times 10^{-7} (2\sigma)$ by analysing the 12.2, 48.372, 48.377 and 60.5-GHz transitions. However, the 12.2-GHz data were likely to have been affected by systematic errors, thus \citet{Kanekar} also completed an analysis with just the high-frequency transitions to obtain a constraint of $\Delta\mu/\mu \leq 4 \times 10^{-7} (2\sigma)$ over a look back time of $\sim 7.5$ Gyr.  

\citet{Simon2011} discuss the major sources of uncertainty which limit the ability of astrophysical observations of methanol to measure (or constrain) changes in $\mu$.  Spatial segregation of molecular species can mimic or hide any true changes in $\mu$.  The excitation conditions differ for each of the transitions, hence at some level the gas responsible for the absorption of each transition will not be identical and hence will not be entirely co-spatial.  The observed Doppler shift of the transition is produced by both turbulence within, and bulk motion of the absorbing gas and must be accounted for in comparing the rest frequencies of the observed transitions.   One advantage of using different transitions of a single molecule to look for changes in $\mu$ is that spatial segregation effects will be much less than when making comparisons between different molecules. However, even for methanol the E and A type rotational transitions essentially correspond to two different species, which can have abundances that differ by up to 40 per cent, depending on the conditions under which the methanol molecules form \citep{Sobolev1997}.  For \pks \citet{Bagd_2013a} found no systematic differences between absorption from 6 different E-type and 4 different A-type methanol transitions, observed with a number of different telescopes covering a frequency range from 6 -- 260~GHz (in the observer frame).  On this basis they ruled out any spatial segregation between the E and A species methanol in \pks and conclude that data from both species should be included in the analysis. This contrasts with a previous analysis by the same authors based on 4 different methanol transitions which, based on the line widths and quality of fit measure, concluded that the two E and A species were spatially segregated \citep{Bagd_2013b}.  Despite these contrasting approaches, it is clear that to minimise potential systematic uncertainties comparison between methanol transitions of the same rotational type and similar upper excitation levels is preferable.

For gravitational lensed systems such as \pks there are a number of additional systematics which have to be considered.  Although gravitational lensing is achromatic, the structure of the background quasar being lensed does vary both with frequency and time.  The quasar jet has a steep spectral index and in \pks it is this which causes an Einstein ring to be present only at lower frequencies \citep{Subrahmanyan+1990}.  In addition, the location at which the core (the base of the jet) becomes optically thin is frequency dependent (so-called core-shift) and this means that different frequencies trace slightly different paths through the lensing galaxy.  For \pks this has been measured at millimetre wavelengths \citep{Marti} and from this we expect core-shifts of the order of tens of microarcseconds at centimetre wavelengths, corresponding to linear scales in the lensing galaxy of around 0.1~pc.  An issue specific to \pks is that the southwestern source has frequency-dependent size, which may cause transitions at different frequencies to trace different paths through the absorbing gas \citep{Kanekar}. For instance, \citet{Kanekar} found that the sightline of the 12.2-GHz transition was likely to be different to that of the higher frequency transitions that it was being compared to. At millimetre wavelengths the absorption in \pks is observed to vary with time due to variability in the structure of the background quasar \citep[e.g.][]{Muller2008}.  This means that to minimise systematic errors observations of different transitions should be taken on timescales less than the timescale of intrinsic variability in the background quasar (of order a month).  Temporal variability in the background quasar will also affect centimetre wavelength transitions, however, the impact can be reduced by observing at very high angular resolution because although the intensity of the most compact emission (the core) may change, its location (and hence line of sight through the lensing galaxy) will not.  An additional advantage of observing at very high angular resolution is that the location of the absorption compared to the continuum emission can be measured, giving a direct indication of the magnitude of the difference in the line of sight.  These systematic sources of uncertainty all need to be considered (and where possible minimised) when using different methanol transitions to search for possible changes in $\mu$. 

Here we report observations of \pks with milliarcsecond angular resolution at 6.45 GHz.  As outlined above, by obtaining high angular resolution data at multiple frequencies at which methanol absorption is observed, the precise locations of absorption features can be used to measure any spatial segregation of different frequency transitions. This can also give information on the effects of the frequency-dependent structure of the source on the path taken through the absorbing gas. 

\section{Observations and Data reduction}
The observations were made using the Long Baseline Array (LBA) on 2013 August 14 (project code V492).  The array for these observations included the Australia Telescope Compact Array (ATCA), Parkes and Mopra antennas operated by the Australia Telescope National Facility and the Hobart and Ceduna telescopes operated by the University of Tasmania. The baseline lengths for this array range from 113 km (ATCA-Mopra) to 1702 km (Hobart-Ceduna).

We used the LBA Data Acquisition System (DAS) to record dual circular polarizations for 2 x 16 MHz bandwidths  with lower-band edge frequencies of 6.434 GHz and 6.450 GHz.  The data were correlated with the DiFX correlator \citep{Deller2011} with 1024 spectral channels for each polarization across the 16 MHz bandwidth (i.e. 15.625 kHz per spectral channel width). The observed frequency range covers the sky frequencies of the 12.2-GHz methanol transition in the northeastern and southwestern components of \pksn, which are redshifted at $z=0.89$ to frequencies of approximately 6.45 GHz. For the 12.2~GHz methanol transition in the rest frame of the absorbing system, this corresponds to a velocity coverage spanning 1485~km~s$^{-1}$ from $-$895 km~s$^{-1}$ to 590 km~s$^{-1}$, with a channel resolution of 0.73~km~s$^{-1}$ and velocity resolution for uniform weighting of the lag function of 0.88~km~s$^{-1}$.  Throughout this paper Doppler-shifted velocities are specified with respect to the barycentric reference frame for an object at redshift $z$=0.88582 (the best estimate of the redshift for the systematic emission from the lensing galaxy).  When comparing our results with those of papers which give velocities with respect to the local standard of rest (LSR) \citep[e.g.][]{Bagd_2013b} subtract 12.432~km~s$^{-1}$ from the LSR velocities to convert them to the barycentric frame for \pksn. 

The primary target for the observations was \pksn, which was observed along with the calibrator sources 3C 273 and PKS\,B1921$-$293 over an observation period of approximately 11 hours.  The pointing center for \pks was $\alpha = 18^h33^m39\fs886$; $\delta = -21^\circ03'40\farcs45$ (J2000) \citep{Subrahmanyan+1990}, which is the location of the southwestern component and this was also the phase centre used for correlation. The total time onsource for \pks was approximately 9 hours and the resulting noise level for the continuum image is 0.7 mJy beam$^{-1}$ (after self-calibration).  Calibrator sources were observed for fringe finding, delay and bandpass calibration purposes.  The absolute flux density scale is estimated to be accurate to 20 per cent.

\begin{figure*}
\begin{center}
	\includegraphics[scale=0.55]{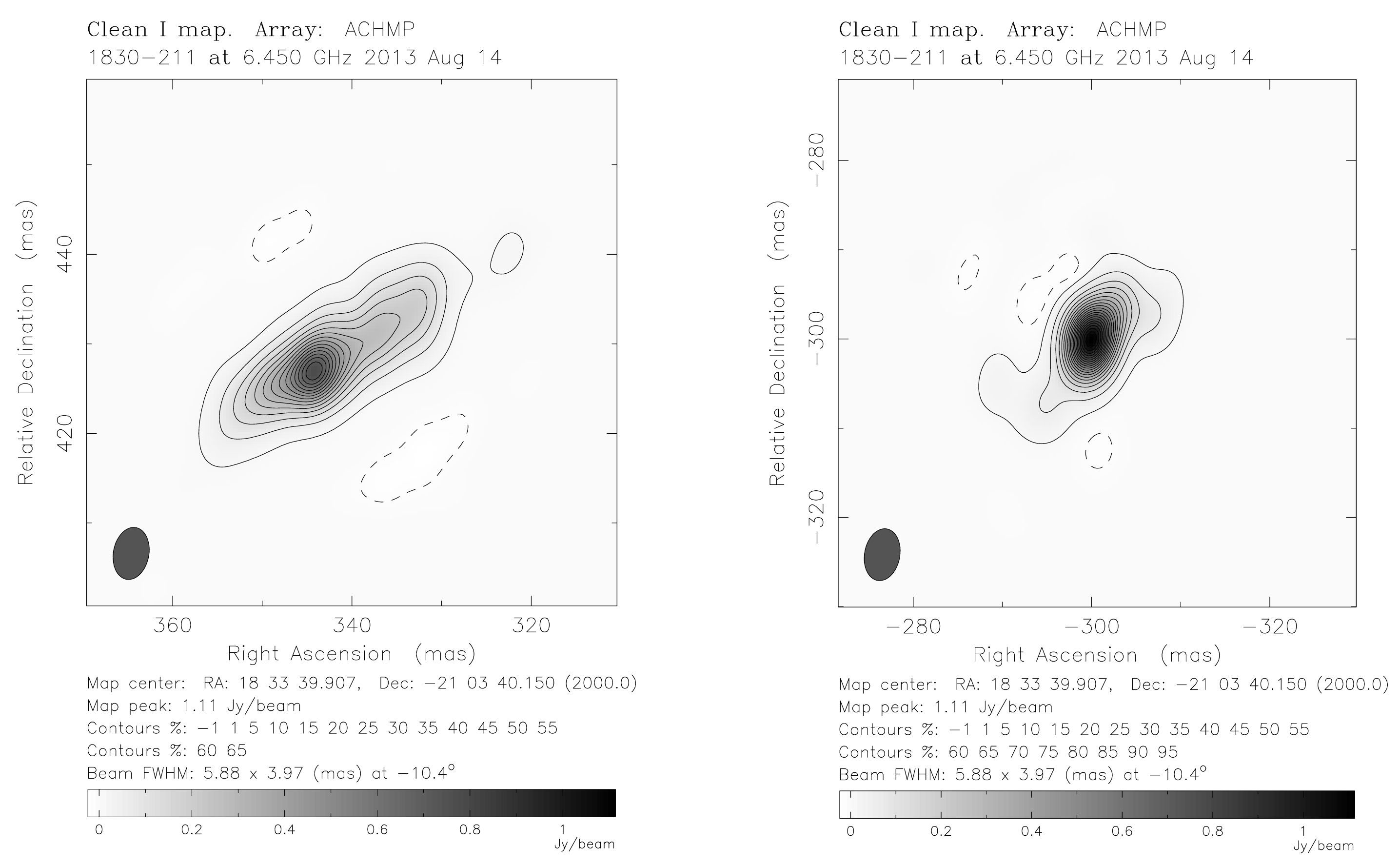}
	\caption{Continuum images of the northeastern \textit{(left)} and southwestern \textit{(right)} components of \pks obtained at 6.45 GHz. The restoring beam is 5.88 by 3.97 mas with the major axis along position angle $-10.4^\circ$ and is shown in the bottom-left corner of each panel. The contours for both panels are at $-1\%$ and 1$\%$, along with each multiple of 5$\%$ up to 95$\%$  of the map peak brightness, which is 1.114 Jy beam$^{-1}$ (the peak of the northeastern component). The RMS noise in these images is 0.7 mJy beam$^{-1}$.}
	\label{allchan}
\end{center}
\end{figure*}

The basic data processing was completed using the Astronomical Image Processing System (AIPS) package, following standard procedures for the reduction of spectral line very long baseline interferometry (VLBI) observations.  Firstly, data over time intervals where the antennas were not onsource was flagged and excluded from the analysis. The amplitudes of the data were then corrected for sampling bias using autocorrelation measurements, a correction which was smoothed with a mean boxcar of width 30 minutes before being applied to the data. The data were then corrected for feed rotation effects by making parallactic angle corrections. Gain factors for each antenna (the square root of the nominal system equivalent flux density) were set in order give approximately correct flux densities for the sources.

In order to calibrate the instrumental delays, a manual phase-calibration fringe fit was run using two minutes of data selected from calibrator source PKS B1921-293, solving for the group delay and phase without solving for fringe rates. The reference antenna used was the ATCA. The phase of the data still varied with frequency, so a further fringe fit was run on the target source \pks over the entire observation interval, with a solution interval of two minutes. The ATCA was again used as the reference antenna. The phases, rates and single-band delays from this fringe fit were smoothed with a median window filter with a width of 5 minutes.  Bandpass calibration was undertaken using PKS B1921-293. The velocity information for \pks was set, with a barycentric velocity system using the optical velocity definition, with line rest frequency of 12.178597 GHz \citep{Muller2004}. Then the data were Doppler corrected so that the mapping of velocity to spectral channels was constant throughout the observations, compensating for the rotation of the antennas with the Earth around the Sun during the observation period.  The data were then amplitude calibrated using the measured system equivalent flux density for each antenna.  The data were exported from AIPS, applying the calibration and excluding edge channels.  

The {\sc difmap} program \citep{Shepherd1994} was used for imaging of the data and self-calibration.  After several iterations of self-calibration of the data using a model of the source based on the clean components from initial imaging, a final image of the source was constructed.  A square map of size 2.9$\times$2.9 arcseconds ($8192\times8192$ pixels with $0.35\times0.35$ milliarcsecond cell size) was created, centred on the approximate mid-point between the two lensed images.  The final continuum image has a synthesised beam size of 5.88$\times$3.97 milliarcseconds at a position angle of $-$10.4 degrees and an RMS noise level of approximately 0.7 mJy beam$^{-1}$.  There is no significant time or bandwidth smearing in the image, due to the relatively high time resolution (5 second samples) and spectral resolution (15.625~kHz) of the imaged data.

\section{Results}

We used the continuum data to create images of the northeastern and southwestern components of \pks and these are shown in Figure \ref{allchan}.   We recover a flux density of 3.73~Jy at 6.45~GHz for this epoch (approximately 90 per cent of the total flux density of the source at the epoch of observation).  Comparing these images with previous VLBI observations \citep[e.g.][]{Guirado1999}, we can see that the general morphology of both the northeastern and southwestern components is similar to that seen at 2.3 and 8.4~GHz.  To facilitate a more quantitative comparison with the results of \citet{Guirado1999} we fitted the strongest component in the northeastern and southwestern components with an elliptical Gaussian.  To estimate the uncertainties in the fitted parameters we took the best-fit model and adjusted a parameter (while holding all others constant), and calculate the reduced chi-squared between the data and perturbed model.  We take the difference between the best fit parameter value and that for which the reduced chi-squared increases by a factor of two, as our estimate of the uncertainty in each parameter. We measured the size of the minor axis of the elliptical Gaussian (which is less affected by source structure than the major axis) to be $2.9\pm1.2$  mas for the northeastern component and $2.0\pm1.2$ mas for the southwestern component.  These are in good agreement with the predicted size at 6.45~GHz \citep[figure 3 of][]{Guirado1999} of approximately 2.6 and 2.0 mas for the northeastern and southwestern components respectively.  We also measured the offset between the peak of the northeastern and southwestern component and find at 6.45 GHz that it is $971.2\pm0.7$~mas at a position angle of $48.6\pm0.1$ degrees, in very good agreement with the 8.4 GHz measurements of \citet{Guirado1999}.

The 12.2-GHz methanol absorption in the southwestern component of \pks is expected to be at a sky frequency 6.45781 GHz for the epoch of the observations \citep[corresponding to a velocity in the rest frame of the absorbing system of $-5.0\pm1.3$ km~s$^{-1}$;][]{Simon2012}. \citet{Simon2012} found that the full-width at half maximum (FWHM) of the absorption is $17.0\pm2.9$ km~s$^{-1}$ for the southwestern component (approximately 20 spectral channels for this data).

In order to further investigate the expected absorption features, a continuum model determined using the data from spectral channels over which no absorption is expected was made and then subtracted from the data. This model was made by following an identical procedure to that taken to obtain the VLBI images in {\sc difmap}, but excluding the approximately 50 spectral channels where we expect absorption (which corresponds to only a few percent of the useful bandwidth of the data). Then this model was set as the continuum model using {\sc difmap}'s \textit{setcont} task. Previous observations suggest that the methanol absorption will be at the level of a few to 10 mJy beam$^{-1}$, which will not be detected with high signal-to-noise in images constructed with a single spectral channel.  To improve the signal-to-noise of any absorption signal, while maintaining modest velocity resolution, we chose to average the data from 9 consecutive spectral channels (velocity width 6.6~km~s$^{-1}$). We then imaged the northeastern and southwestern components with the continuum subtracted data, using the 9 spectral channels centred on the peak of the two absorption components.  We also imaged two other 9 channel ranges in which no absorption is expected, for comparison.  We detected absorption at the expected velocity towards the southwestern component, but not in any of the other images.  Some of the other images do show large positive and negative features, however, none are projected against the continuum emission.  The relatively poor quality of the continuum subtracted images is primarily due to the limited dynamic range of the final image cube, which we discuss in more detail below. To test the robustness of the detection of absorption towards the southwestern component we made additional images covering 2 ranges of 9 spectral channel each, on either side of the centre of the absorption peak (i.e. each of the two ranges contained 4 or 5 channels from the original data and 4 or 5 channels further from the centre of the absorption peak).  This same absorption feature was detected clearly in both datasets at the same location, providing further evidence that we are indeed detecting the methanol absorption from the southwestern source component.  Any absorption towards the northeastern component has a depth of less than 5 mJy beam$^{-1}$.

\begin{figure}
\begin{center}
	\includegraphics[scale=0.45]{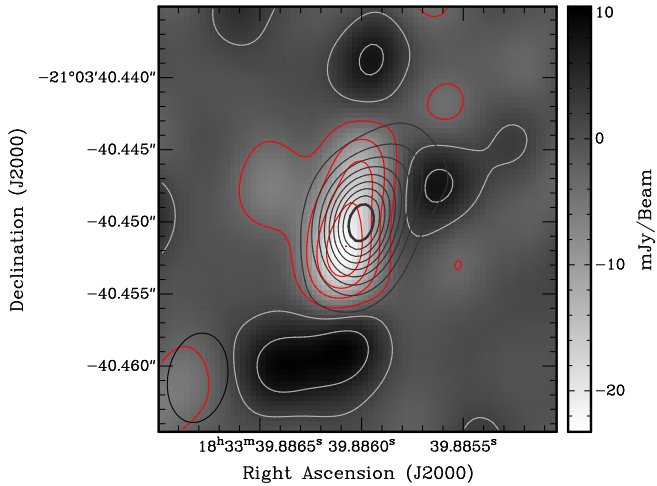} 
	\caption{The greyscale image shows the southwestern component of the continuum subtracted data from the spectral channels centred on the expected southwestern absorption, with a channel range with width equal to the expected FWHM of the absorption ($\sim$ 17~km~s$^{-1}$ or 23 spectral channels). The absorption has depth 23.2 mJy~beam$^{-1}$, with RMS noise in the image of 1.4~mJy~beam$^{-1}$. The red contours show the absorption in steps of 15, 30, 45 and 60 per cent of the peak absorption, the light grey contours are the positive contours from the continuum subtracted data at the same percentage levels. The black contours show the continuum emission from the southwestern continuum component with levels every 10 per cent of the peak between 10 and 90 percent. The black ellipse in the bottom-left corner represents the synthesised beam.}
	\label{FWHM}
\end{center}
\end{figure}

To obtain the best possible image of the southwestern absorption from the current data we then constructed an image from the continuum subtracted data averaged over the $\sim$ 17~km~s$^{-1}$ (23 spectral channels) centred on the frequency of the southwestern absorption component.  This corresponds to the FWHM of the absorption observed with the Australia Telescope Compact Array at this frequency \citep{Simon2012}.  Figure~\ref{FWHM} shows an image of the absorption towards the southwestern component superimposed on the continuum emission.  The peak of the absorption ($\sim$ $-$23 mJy beam$^{-1}$) is a factor of 17 greater than the RMS noise level in the image (1.4 mJy beam$^{-1}$) and is projected against the continuum emission.  The peak of the absorption is located to the south-east of the peak of the continuum emission, at $\alpha = 18^h33^m39\fs88608$; $\delta = -21^\circ03'40\farcs4517$ (J2000).  The peak of the continuum emission is measured to be at $\alpha = 18^h33^m39\fs88600$; $\delta = -21^\circ03'40\farcs4500$ (J2000), the phase centre, as expected for self-calibrated data.  For the adopted cosmology 1 arcsecond corresponds to a linear scale of 8.04 kpc, so the angular offset of 2.0 milliarcseconds between the peaks in the continuum and absorption (approximately 40 per cent of the synthesised beam width), corresponds to a linear offset of approximately 16 pc. Taking slices through the major and minor axes of the absorption image we estimate a size of $6.3 \times 4.2$ mas at a position angle of $80^{\circ}$, slightly larger than the synthesised beam.  We also fitted an elliptical Gaussian and a point source model to the absorption and found that although the elliptical Gaussian gives a lower reduced chi-squared for the fit, it is only marginally better than for a point source.  The limited signal-to-noise in the absorption image means that although there is some evidence that the absorption is partly resolved we cannot be certain of this from the image alone. 

To test the robustness of the observed offset of the maximum absorption from the peak of the continuum emission we produced continuum images with similar properties to the  absorption dataset.  The first test we undertook was to take the final clean model of the continuum dataset and subtract 98 per cent of each clean component from the uv-data.  The result is a dataset with 2 per cent of the original continuum emission, which is equivalent to the peak absorption depth observed and has similar noise characteristics.  We then selected 23 consecutive spectral channels and imaged the emission using the same procedure as applied to the absorption imaging and measured the location of the peak of the south-western continuum component.  This process was repeated a further 24 times (25 iterations in total) and the RMS offset of the south-western continuum peak from the position determined from the best continuum image was measured to be 0.71 mas.  As a second test we created an additional uv-dataset with 2 per cent of the total continuum flux density.  Rather than reducing all clean components by the same fraction, we selected a clean component at random, reduced its amplitude by a random amount which varied between 1 and 3 per cent and repeated this process until the total flux density summed over all clean components was 98 per cent of the original total flux density.  This model was then subtracted from the uv-data resulting in a dataset with 2 per cent of the original continuum emission.  We then repeated the first test by imaging 25 different combinations of 23 consecutive spectral channels and determining the location of the peak of the continuum emission for the south-western component.  The RMS offset measured for this second test was 0.77 mas.  If we take the average of these two tests as representative of the uncertainty in the position measured in the absorption image (0.74 mas), then the absorption is offset from the continuum peak by 2.7-$\sigma$.  Since the strength of the absorption depends on the gas column density in the lensing galaxy along the line of sight to the SW (or NE) component and this is expected have structure on scales of 1-10 pc (corresponding to angular scales of  0.1-1 mas), it is more likely than not that the strongest absorption will be offset from the strongest emission.  Our relative astrometric uncertainty does not allow us to definitively demonstrate this, but it is consistent with there being an offset.
  
\begin{figure}
\begin{center}
      \includegraphics[scale=0.4,angle=270]{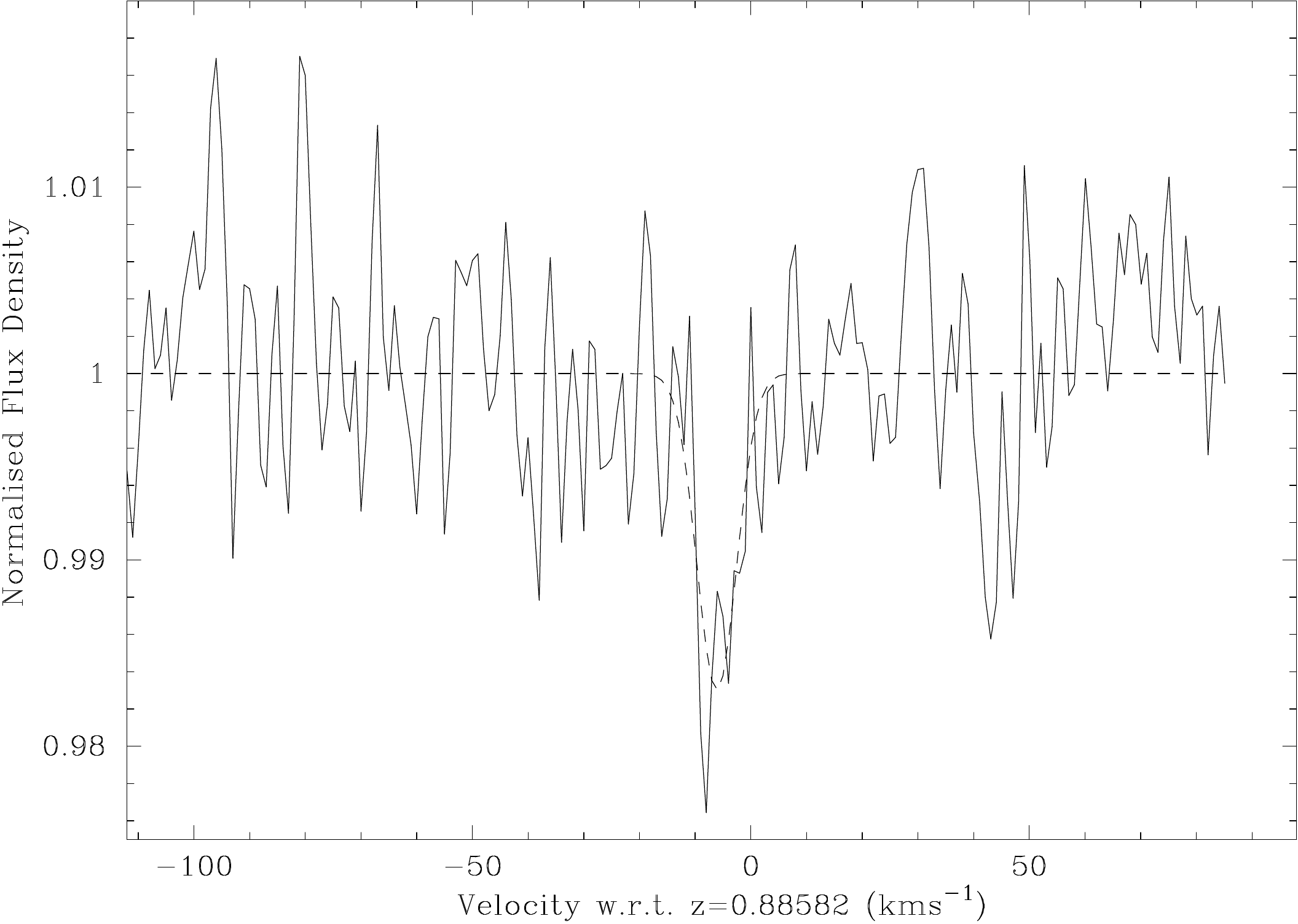}
\end{center}
\caption{The absorption spectrum in the southwestern component from the LBA data using the data from all baselines, with 1~km~s$^{-1}$ spectral channels.  The plot shows the barycentric velocity of the 12.2~GHz methanol transition with respect to $z$ = 0.88582 versus the normalised flux density.  The dashed line is a Gaussian fit to the absorption which has a peak optical depth $\tau$ = 0.017, with a FWHM of 8.4~km~s$^{-1}$ centred at a barycentric velocity of $-6.1$~km~s$^{-1}$}  \label{fig:abs}
\end{figure}

We were not able to produce vector-averaged spectra of the self-calibrated image cube data in {\sc difmap}, so we exported the dataset to {\sc miriad} for further analysis.  The final data reduction of the $(u,v)$ data prior to continuum subtraction involved two passes.  For each pass, the data were phase-shifted so that first the northeastern and then the southwestern continuum component was at the phase centre.  In each case, simple vector averaging ({\sc miriad} task {\sc uvspec}) gives a good estimate of the spectrum toward the new map centre, as measured on each baseline.  Working on the uv-plane in this way eliminates effects due to convolving the data onto a uv grid, Fourier inversion to the map plane, and deconvolving the effect of the resulting dirty beam.  The spectrum for each baseline of the array was averaged over time, and then the spectra from all baselines were averaged.  Figure~\ref{fig:abs} shows a vector-averaged spectrum with 1~km~s$^{-1}$ spectral channels constructed using data from all baselines and with the phase-centre set to the location of the southwestern continuum peak.   Fitting a Gaussian to the absorption we find an optical depth $\tau = 0.017\pm0.003$ at a barycentric velocity of $-6.1\pm0.7$~km~s$^{-1}$, the velocity expected for absorption from the southwestern methanol component.  The FWHM of the absorption component is $8.4\pm1.6$~km~s$^{-1}$.  The continuum emission of the southwestern component is 1.11 Jy at this frequency, so an optical depth of 0.017 corresponds to $19\pm3$~mJy, which is consistent with the absorption detected in the continuum subtracted image data (Fig.~\ref{FWHM}).

\section{Discussion}

Our observations of the 12.2-GHz methanol absorption towards the southwestern component of PKS\,B1830$-$211 with an angular resolution of around 5 milliarcseconds are significantly higher resolution than any previous observations of the molecular absorption in this source.  For the adopted cosmology ($\textrm{H}_0 = 67$ km~s$^{-1}$ Mpc$^{-1}$, $\Omega_m = 0.315$ and $\Omega_\Lambda = 0.685$) this angular resolution corresponds to a linear scale of 40~pc at $z$=0.88582 \citep{Wright2006}, approximately the same scale size as an individual giant molecular cloud \citep{Blitz1980}.  Previous observations of the methanol absorption with angular resolutions of the order of arcseconds to arcminutes show absorption with a maximum at the same velocity, but lower optical depth and greater line width \citep{Simon2012,Bagd_2013a,Bagd_2013b,Kanekar}.  The highest signal-to-noise spectrum of the 12.2-GHz methanol absorption was obtained with the Very Large Array (VLA) by \citet{Kanekar}, who find it to have an optical depth of approximately 0.0026 and a FWHM greater than 13.4~km~s$^{-1}$.  The VLA observations have an angular resolution of around 1 arcsecond and do not resolve the northeastern and southwestern components of PKS\,B1830$-$211, nor the Einstein ring observed at frequencies less than 10~GHz \citep{Jauncey}.  \citet{Kanekar} find a weak absorption wing at positive velocities in their spectrum, which is not seen in the higher frequency methanol transitions they observed.  From this they infer the presence of multiple absorbing components in the 12.2~GHz spectrum, some of which are not present in the other transitions.


\citet{Simon2012} marginally detected 12.2-GHz absorption at a velocity of $-125$~km~s$^{-1}$, which they claimed may be absorption towards the northeastern component of PKS\,B1830$-$211.  The 12.2-GHz absorption spectrum shown by \citet{Kanekar} does not cover this velocity range and the highest signal-to-noise spectrum of the 12.2-GHz absorption presented in \citet{Bagd_2013b} (obtained with the Effelsberg 100-m) does not show this component.  The current observations also do not detect any absorption from the northeastern component at levels comparable to the marginal detection of \citet{Simon2012}, which suggests that it is likely not real. 

Our observations with angular resolutions of order a few milliarcseconds are able to resolve individual giant molecular clouds in the lensing galaxy of the PKS\,B1830$-$211 system.  The most comprehensive study of 12.2-GHz methanol absorption in the Milky Way is that of \citet{Peng1992}, who observed with the Parkes telescope with 2.1~arcminute angular resolution, which corresponds to approximately a factor of 50 higher linear resolution than we achieve for \pksn.  They found a median linewidth of 4.2~km~s$^{-1}$, although in a number of cases linewidths in excess of 10~km~s$^{-1}$ were observed for sources well outside the Galactic Centre region.  So the 12.2-GHz methanol absorption detected in \pks at milliarcsecond resolution likely arises from one or two molecular cores in the lensing galaxy.  

The optical depth of the 12.2-GHz absorption on milliarcsecond scales is an order of magnitude greater than that observed at angular resolutions of an arcsecond and larger and is comparable to that observed in the higher frequency transitions \citep{Bagd_2013a,Kanekar}.  The width of the absorption we detect is approximately 30 per cent narrower than that seen in any previous observations of this transition.  There are 5 observations of the $2_0 \rightarrow 3_{-1} E$ transition of methanol towards \pks reported in the published literature, with data collected between 2011 November and 2013 May \citep{Simon2012,Bagd_2013a,Bagd_2013b,Kanekar}.  The width of the absorption line measured from these observations varies between 12 and 20~km~s$^{-1}$ and the velocity of the line centre between -5 and 0~km~s$^{-1}$ (in the barycentric reference frame for $z$=0.88582).  Variability in the profile of absorption lines observed towards \pks can potentially arise either through changes in the morphology of the background quasar, or from the relative motion of molecular clouds in the lensing galaxy.  Significant variability in absorption line profiles for \pks have previously been observed in a number of other transitions \citep[e.g.][]{Muller2008}, predominantly in lower-excitation lines at millimetre wavelengths such as HCO$^+$ and CS.  For some of the millimetre wavelength lines there is good evidence that the molecular emission completely covers the SW continuum source \citep[e.g.][]{Muller2014} and where significant changes in the absorption profile are observed the evidence suggests that it is predominantly due to variability in the background quasar \citep[e.g.][]{Schulz}, which is significant on short timescales at millimetre wavelengths \citep{Marti}.  The $2_0 \rightarrow 3_{-1} E$ methanol transition shows much narrower and weaker absorption, indicating optically thin absorption and a much lower covering factor.  The lower frequency of this transition means that the emission from the background quasar covers a larger angular scale and shows less rapid variability than that at higher frequencies.  These two factors mean that changes in the absorption profile of the 12.2~GHz methanol transition are expected to be both smaller and slower than those which are observed for millimetre wavelength transitions.  The current observations measure a full-width half maximum for the $2_0 \rightarrow 3_{-1} E$ line of $8.4\pm1.6$~km~s$^{-1}$, suggesting that some of the absorption is resolved at milliarcsecond scales.  Without contemporaneous observations at lower angular resolution for comparison we cannot demonstrate conclusively that we are resolving some of the absorption, however, given that observations with the Efflesberg 100-m telescope three months prior to our LBA observations measured a line width of $19.9\pm4.5$~km~s$^{-1}$, it is highly likely that we are.

The angular size of the southwestern core component in \pks is known to vary with frequency \citep{Guirado1999}, however, the increased similarity of the optical depth and line profile for the 12.2-GHz transition to the higher frequency transitions when observed at milliarcsecond resolution suggests that it is possible to significantly reduce sources of systematic error, or at least improve our ability to estimate them.   \citet{Marti} estimate the scale of the frequency-dependent core shift in \pks at the frequencies of the methanol absorption measurements to be around 0.1 mas, an order of magnitude less than the offset we observe between the peak of the continuum emission and the strongest absorption.  If the angular scale of the primary absorption system is significantly larger than 0.1 mas (which appears likely from our observations), then this reduces the impact of core shift on uncertainty in measurements of $\Delta\mu/\mu$. Hence, combining data from the 12.2-GHz absorption system with similar high resolution observations of other methanol transitions towards PKS\,B1830$-$211 will enable significant improvements in constraints in $\Delta\mu/\mu$.  This is important, as \cite{Thom2012} shows that existing constraints on $\Delta\mu/\mu$ are very difficult to reconcile theoretically with claims of observed changes in the fine structure constant $\alpha$ \citep{Webb2011}.  When considering observations of absorption from two different methanol transitions, the constraint that this provides is:
\begin{equation}
\frac{\Delta\mu}{\mu} \le \frac{\Delta v}{\Delta K_{\mu} c} \label{eqn:calc}
\end{equation}
where $c$ is the speed of light, $\Delta v$ is the difference in the measured velocity of the absorption and $\Delta K_{\mu}$ is the difference in the sensitivity coefficients.  This paper reports milliaracsecond resolution observations of the 12.2-GHz methanol transition, however, there relatively few other detected transitions accessible to centimetre wavelength VLBI arrays.  The 48.3-GHz transitions (redshifted to $\sim$25.6~GHz) lie outside the frequency range covered by the VLBA, but can potentially be observed using the LBA.  These transitions have $K_{\mu}$~=~-1 (compared to -33 for the 12.2-GHz transition), so from Equation~\ref{eqn:calc}, a 1~km~s$^{-1}$ uncertainty in the velocity difference of the absorption lines from these transitions corresponds to a limit in $\Delta{\mu}/\mu$ of $1 \times 10^{-7}$, comparable to the best limits achieved to date \citep[e.g.][]{Bagd_2013a,Kanekar}.  The current observations provide an estimate of the velocity of the maximum absorption of the 12.2-GHz transition with a 1$\sigma$ uncertainty of around 1~km~s$^{-1}$.  Observations of the 12.2- and 48.3-GHz transitions with sufficient signal-to-noise to measure the velocity difference of the maximum absorption with an uncertainty less than 0.53~km~s$^{-1}$ are required to obtain a constraint better than that claimed by \citet{Kanekar}.  However, provided the observations are made sufficiently close together to nullify temporal variation in the absorption system, then potential systematic error sources are minimised, and it could be argued these have been underestimated in some previous investigations using arcsecond resolution data.  Furthermore, future high resolution ALMA observations may be able to undertake similar investigations with marginally lower resolution (around 20 milliarcseconds), but for many more transitions and at higher signal-to-noise than can likely be achieved using centimetre wavelength VLBI.

The structure of PKS\,B1830$-$211, with two compact components of similar intensity separated by approximately 1 arcsecond presents some significant challenges for imaging with a hetrogeneous VLBI array with relatively few elements such as the LBA.  The relatively sparse $(u,v)$-coverage limits the maximum dynamic range that the LBA can achieve in imaging the emission in such a source.  This in turn limits the accuracy of the model for continuum subtraction, introducing large-scale baseline-dependent ripples in the spectrum which reduces our ability to detect weak and broad absorption features in the source.  The Very Long Baseline Array (VLBA) has a large number of antennas (10, compared to the 5 antennas in the array for these observations) and as they are identical, calibration is generally easier than for a hetrogeneous array.  Hence observations of the 12.2-GHz transition in PKS\,B1830$-$211 with the VLBA with similar sensitivity, should be able to achieve superior data on the absorption system.  Unfortunately the 48.3- and 60.5-GHz methanol transitions for PKS\,B1830$-$211 correspond to observing frequencies of approximately 25.6 and 30.2~GHz, which are outside the range of existing receivers on the VLBA.

\section{Conclusion}
We have imaged the 12.2-GHz methanol absorption from the southwestern component of PKS\,B1830$-$211 at an angular resolution of 5 milliarcseconds, corresponding to a linear scale of 40~pc in the $z$=0.88582 galaxy.  This is the first time that molecular absorption in PKS\,B1830$-$211 has been imaged at resolution sufficient to partially resolve it.  The detection of absorption on these scales demonstrates that it is possible to improve on the best current constraints on $\Delta \mu/\mu$ that have been obtained in this source by making similar resolution observations of other methanol transitions.  Observed astronomical constraints on $\mu$ and $\alpha$ already provide meaningful constraints on cosmological models and new physics \citep{Thom2012}, so potential improvements in the sensitivity of $\Delta \mu/\mu$ observations have broad implications for a variety of branches of physics.

\bibliography{LitReviewBib}

\begin{thebibliography}{}
\makeatletter
\relax
\def\mn@urlcharsother{\let\do\@makeother \do\$\do\&\do\#\do\^\do\_\do\%\do\~}
\def\mn@doi{\begingroup\mn@urlcharsother \@ifnextchar [ {\mn@doi@}
  {\mn@doi@[]}}
\def\mn@doi@[#1]#2{\def\@tempa{#1}\ifx\@tempa\@empty \href
  {http://dx.doi.org/#2} {doi:#2}\else \href {http://dx.doi.org/#2} {#1}\fi
  \endgroup}
\def\mn@eprint#1#2{\mn@eprint@#1:#2::\@nil}
\def\mn@eprint@arXiv#1{\href {http://arxiv.org/abs/#1} {{\tt arXiv:#1}}}
\def\mn@eprint@dblp#1{\href {http://dblp.uni-trier.de/rec/bibtex/#1.xml}
  {dblp:#1}}
\def\mn@eprint@#1:#2:#3:#4\@nil{\def\@tempa {#1}\def\@tempb {#2}\def\@tempc
  {#3}\ifx \@tempc \@empty \let \@tempc \@tempb \let \@tempb \@tempa \fi \ifx
  \@tempb \@empty \def\@tempb {arXiv}\fi \@ifundefined
  {mn@eprint@\@tempb}{\@tempb:\@tempc}{\expandafter \expandafter \csname
  mn@eprint@\@tempb\endcsname \expandafter{\@tempc}}}

\bibitem[\protect\citeauthoryear{{Bagdonaite}, {Dapr{\`a}}, {Jansen},
  {Bethlem}, {Ubachs}, {Muller}, {Henkel}  \& {Menten}}{{Bagdonaite}
  et~al.}{2013a}]{Bagd_2013a}
{Bagdonaite} J.,  {Dapr{\`a}} M.,  {Jansen} P.,  {Bethlem} H.~L.,  {Ubachs} W.,
   {Muller} S.,  {Henkel} C.,   {Menten} K.~M.,  2013a, \mn@doi [Physical
  Review Letters] {10.1103/PhysRevLett.111.231101}, \href
  {http://adsabs.harvard.edu/abs/2013PhRvL.111w1101B} {111, 231101}

\bibitem[\protect\citeauthoryear{{Bagdonaite}, {Jansen}, {Henkel}, {Bethlem},
  {Menten}  \& {Ubachs}}{{Bagdonaite} et~al.}{2013b}]{Bagd_2013b}
{Bagdonaite} J.,  {Jansen} P.,  {Henkel} C.,  {Bethlem} H.~L.,  {Menten} K.~M.,
    {Ubachs} W.,  2013b, \mn@doi [Science] {10.1126/science.1224898}, \href
  {http://adsabs.harvard.edu/abs/2013Sci...339...46B} {339, 46}

\bibitem[\protect\citeauthoryear{{Blitz} \& {Shu}}{{Blitz} \&
  {Shu}}{1980}]{Blitz1980}
{Blitz} L.,  {Shu} F.~H.,  1980, \mn@doi [\apj] {10.1086/157968}, \href
  {http://adsabs.harvard.edu/abs/1980ApJ...238..148B} {238, 148}

\bibitem[\protect\citeauthoryear{{Deller} et~al.,}{{Deller}
  et~al.}{2011}]{Deller2011}
{Deller} A.~T.,  et~al., 2011, \mn@doi [\pasp] {10.1086/658907}, \href
  {http://adsabs.harvard.edu/abs/2011PASP..123..275D} {123, 275}

\bibitem[\protect\citeauthoryear{{Ellingsen}, {Voronkov}  \&
  {Breen}}{{Ellingsen} et~al.}{2011}]{Simon2011}
{Ellingsen} S.,  {Voronkov} M.,   {Breen} S.,  2011, \mn@doi [Physical Review
  Letters] {10.1103/PhysRevLett.107.270801}, \href
  {http://adsabs.harvard.edu/abs/2011PhRvL.107A0801E} {107, 270801}

\bibitem[\protect\citeauthoryear{{Ellingsen}, {Voronkov}, {Breen}  \&
  {Lovell}}{{Ellingsen} et~al.}{2012}]{Simon2012}
{Ellingsen} S.~P.,  {Voronkov} M.~A.,  {Breen} S.~L.,   {Lovell} J.~E.~J.,
  2012, \mn@doi [\apjl] {10.1088/2041-8205/747/1/L7}, \href
  {http://adsabs.harvard.edu/abs/2012ApJ...747L...7E} {747, L7}

\bibitem[\protect\citeauthoryear{{Guirado}, {Jones}, {Lara}, {Marcaide},
  {Preston}, {Rao}  \& {Sherwood}}{{Guirado} et~al.}{1999}]{Guirado1999}
{Guirado} J.~C.,  {Jones} D.~L.,  {Lara} L.,  {Marcaide} J.~M.,  {Preston}
  R.~A.,  {Rao} A.~P.,   {Sherwood} W.~A.,  1999, \aap, \href
  {http://adsabs.harvard.edu/abs/1999A%26A...346..392G} {346, 392}

\bibitem[\protect\citeauthoryear{{Jansen}, {Kleiner}, {Xu}, {Ubachs}  \&
  {Bethlem}}{{Jansen} et~al.}{2011}]{Jansen}
{Jansen} P.,  {Kleiner} I.,  {Xu} L.-H.,  {Ubachs} W.,   {Bethlem} H.~L.,
  2011, \mn@doi [\pra] {10.1103/PhysRevA.84.062505}, \href
  {http://adsabs.harvard.edu/abs/2011PhRvA..84f2505J} {84, 062505}

\bibitem[\protect\citeauthoryear{{Jauncey} et~al.,}{{Jauncey}
  et~al.}{1991}]{Jauncey}
{Jauncey} D.~L.,  et~al., 1991, \mn@doi [\nat] {10.1038/352132a0}, \href
  {http://adsabs.harvard.edu/abs/1991Natur.352..132J} {352, 132}

\bibitem[\protect\citeauthoryear{{Kanekar} et~al.,}{{Kanekar}
  et~al.}{2015}]{Kanekar}
{Kanekar} N.,  et~al., 2015, \mn@doi [\mnras] {10.1093/mnrasl/slu206}, \href
  {http://adsabs.harvard.edu/abs/2015MNRAS.448L.104K} {448, L104}

\bibitem[\protect\citeauthoryear{{Levshakov}, {Kozlov}  \&
  {Reimers}}{{Levshakov} et~al.}{2011}]{LevSept}
{Levshakov} S.~A.,  {Kozlov} M.~G.,   {Reimers} D.,  2011, \mn@doi [\apj]
  {10.1088/0004-637X/738/1/26}, \href
  {http://adsabs.harvard.edu/abs/2011ApJ...738...26L} {738, 26}

\bibitem[\protect\citeauthoryear{{Lidman}, {Courbin}, {Meylan}, {Broadhurst},
  {Frye}  \& {Welch}}{{Lidman} et~al.}{1999}]{Lidman}
{Lidman} C.,  {Courbin} F.,  {Meylan} G.,  {Broadhurst} T.,  {Frye} B.,
  {Welch} W.~J.~W.,  1999, \mn@doi [\apjl] {10.1086/311949}, \href
  {http://adsabs.harvard.edu/abs/1999ApJ...514L..57L} {514, L57}

\bibitem[\protect\citeauthoryear{{Lovell} et~al.,}{{Lovell}
  et~al.}{1996}]{Lovell}
{Lovell} J.~E.~J.,  et~al., 1996, \mn@doi [\apjl] {10.1086/310353}, \href
  {http://adsabs.harvard.edu/abs/1996ApJ...472L...5L} {472, L5}

\bibitem[\protect\citeauthoryear{{Lovell}, {Jauncey}, {Reynolds}, {Wieringa},
  {King}, {Tzioumis}, {McCulloch}  \& {Edwards}}{{Lovell}
  et~al.}{1998}]{Lovell+98}
{Lovell} J.~E.~J.,  {Jauncey} D.~L.,  {Reynolds} J.~E.,  {Wieringa} M.~H.,
  {King} E.~A.,  {Tzioumis} A.~K.,  {McCulloch} P.~M.,   {Edwards} P.~G.,
  1998, \mn@doi [\apjl] {10.1086/311723}, \href
  {http://adsabs.harvard.edu/abs/1998ApJ...508L..51L} {508, L51}

\bibitem[\protect\citeauthoryear{{Mart{\'{\i}}-Vidal}
  et~al.,}{{Mart{\'{\i}}-Vidal} et~al.}{2013}]{Marti}
{Mart{\'{\i}}-Vidal} I.,  et~al., 2013, \mn@doi [\aap]
  {10.1051/0004-6361/201322131}, \href
  {http://adsabs.harvard.edu/abs/2013A%26A...558A.123M} {558, A123}

\bibitem[\protect\citeauthoryear{{Muller} \& {Gu{\'e}lin}}{{Muller} \&
  {Gu{\'e}lin}}{2008}]{Muller2008}
{Muller} S.,  {Gu{\'e}lin} M.,  2008, \mn@doi [\aap]
  {10.1051/0004-6361:200810392}, \href
  {http://adsabs.harvard.edu/abs/2008A%26A...491..739M} {491, 739}

\bibitem[\protect\citeauthoryear{{M{\"u}ller}, {Menten}  \&
  {M{\"a}der}}{{M{\"u}ller} et~al.}{2004}]{Muller2004}
{M{\"u}ller} H.~S.~P.,  {Menten} K.~M.,   {M{\"a}der} H.,  2004, \mn@doi [\aap]
  {10.1051/0004-6361:20041384}, \href
  {http://adsabs.harvard.edu/abs/2004A%26A...428.1019M} {428, 1019}

\bibitem[\protect\citeauthoryear{{Muller}, {Gu{\'e}lin}, {Dumke}, {Lucas}  \&
  {Combes}}{{Muller} et~al.}{2006}]{Muller2006}
{Muller} S.,  {Gu{\'e}lin} M.,  {Dumke} M.,  {Lucas} R.,   {Combes} F.,  2006,
  \mn@doi [\aap] {10.1051/0004-6361:20065187}, \href
  {http://adsabs.harvard.edu/abs/2006A%26A...458..417M} {458, 417}

\bibitem[\protect\citeauthoryear{{Muller} et~al.,}{{Muller}
  et~al.}{2011}]{Muller2011}
{Muller} S.,  et~al., 2011, \mn@doi [\aap] {10.1051/0004-6361/201117096}, \href
  {http://adsabs.harvard.edu/abs/2011A%26A...535A.103M} {535, A103}

\bibitem[\protect\citeauthoryear{{Muller} et~al.,}{{Muller}
  et~al.}{2014}]{Muller2014}
{Muller} S.,  et~al., 2014, \mn@doi [\aap] {10.1051/0004-6361/201423646}, \href
  {http://adsabs.harvard.edu/abs/2014A%26A...566A.112M} {566, A112}

\bibitem[\protect\citeauthoryear{{Peng} \& {Whiteoak}}{{Peng} \&
  {Whiteoak}}{1992}]{Peng1992}
{Peng} R.~S.,  {Whiteoak} J.~B.,  1992, \mn@doi [\mnras]
  {10.1093/mnras/254.2.301}, \href
  {http://adsabs.harvard.edu/abs/1992MNRAS.254..301P} {254, 301}

\bibitem[\protect\citeauthoryear{{Planck Collaboration} et~al.,}{{Planck
  Collaboration} et~al.}{2014}]{Planck}
{Planck Collaboration} et~al., 2014, \mn@doi [\aap]
  {10.1051/0004-6361/201321591}, \href
  {http://adsabs.harvard.edu/abs/2014A%26A...571A..16P} {571, A16}

\bibitem[\protect\citeauthoryear{{Schulz}, {Henkel}, {Menten}, {Muller},
  {Muders}, {Bagdonaite}  \& {Ubachs}}{{Schulz} et~al.}{2015}]{Schulz}
{Schulz} A.,  {Henkel} C.,  {Menten} K.~M.,  {Muller} S.,  {Muders} D.,
  {Bagdonaite} J.,   {Ubachs} W.,  2015, \mn@doi [\aap]
  {10.1051/0004-6361/201425072}, \href
  {http://adsabs.harvard.edu/abs/2015A%26A...574A.108S} {574, A108}

\bibitem[\protect\citeauthoryear{{Shepherd}, {Pearson}  \& {Taylor}}{{Shepherd}
  et~al.}{1994}]{Shepherd1994}
{Shepherd} M.~C.,  {Pearson} T.~J.,   {Taylor} G.~B.,  1994, in Bulletin of the
  American Astronomical Society. pp 987--989

\bibitem[\protect\citeauthoryear{{Sobolev}, {Cragg}  \& {Godfrey}}{{Sobolev}
  et~al.}{1997}]{Sobolev1997}
{Sobolev} A.~M.,  {Cragg} D.~M.,   {Godfrey} P.~D.,  1997, \aap, \href
  {http://adsabs.harvard.edu/abs/1997A%26A...324..211S} {324, 211}

\bibitem[\protect\citeauthoryear{{Subrahmanyan}, {Narasimha}, {Pramesh-Rao}  \&
  {Swarup}}{{Subrahmanyan} et~al.}{1990}]{Subrahmanyan+1990}
{Subrahmanyan} R.,  {Narasimha} D.,  {Pramesh-Rao} A.,   {Swarup} G.,  1990,
  \mnras, \href {http://adsabs.harvard.edu/abs/1990MNRAS.246..263S} {246, 263}

\bibitem[\protect\citeauthoryear{{Thompson}}{{Thompson}}{2012}]{Thom2012}
{Thompson} R.~I.,  2012, \mn@doi [\mnras] {10.1111/j.1745-3933.2012.01238.x},
  \href {http://adsabs.harvard.edu/abs/2012MNRAS.422L..67T} {422, L67}

\bibitem[\protect\citeauthoryear{{Uzan}}{{Uzan}}{2011}]{Uzan}
{Uzan} J.-P.,  2011, \mn@doi [Living Reviews in Relativity]
  {10.12942/lrr-2011-2}, \href
  {http://adsabs.harvard.edu/abs/2011LRR....14....2U} {14, 2}

\bibitem[\protect\citeauthoryear{{Varshalovich} \& {Levshakov}}{{Varshalovich}
  \& {Levshakov}}{1993}]{Varsha}
{Varshalovich} D.~A.,  {Levshakov} S.~A.,  1993, Soviet Journal of Experimental
  and Theoretical Physics Letters, \href
  {http://adsabs.harvard.edu/abs/1993JETPL..58..237V} {58, 237}

\bibitem[\protect\citeauthoryear{{Webb}, {King}, {Murphy}, {Flambaum},
  {Carswell}  \& {Bainbridge}}{{Webb} et~al.}{2011}]{Webb2011}
{Webb} J.~K.,  {King} J.~A.,  {Murphy} M.~T.,  {Flambaum} V.~V.,  {Carswell}
  R.~F.,   {Bainbridge} M.~B.,  2011, \mn@doi [Physical Review Letters]
  {10.1103/PhysRevLett.107.191101}, \href
  {http://adsabs.harvard.edu/abs/2011PhRvL.107s1101W} {107, 191101}

\bibitem[\protect\citeauthoryear{{Wiklind} \& {Combes}}{{Wiklind} \&
  {Combes}}{1996}]{Wiklind}
{Wiklind} T.,  {Combes} F.,  1996, \mn@doi [\nat] {10.1038/379139a0}, \href
  {http://adsabs.harvard.edu/abs/1996Natur.379..139W} {379, 139}

\bibitem[\protect\citeauthoryear{{Wright}}{{Wright}}{2006}]{Wright2006}
{Wright} E.~L.,  2006, \mn@doi [\pasp] {10.1086/510102}, \href
  {http://adsabs.harvard.edu/abs/2006PASP..118.1711W} {118, 1711}

\makeatother
\end{thebibliography}

\label{lastpage}

\end{document}